\documentclass[a4paper,english]{article}
\usepackage{fontenc}
\usepackage[latin1]{inputenc}
\usepackage{amsmath}
\usepackage{graphicx}
\usepackage{amssymb}
\usepackage{fontenc}
\usepackage{amsmath}
\usepackage{graphicx}
\usepackage{amssymb}
\usepackage{graphicx}
\usepackage{amssymb}
\usepackage{graphicx}
\usepackage{amssymb}
\usepackage{graphicx}
\usepackage{amssymb}
\usepackage{amssymb}
\usepackage{babel}
\usepackage{babel}
\usepackage{babel}
\usepackage{babel}

\input{tcilatex}
\input psfig.sty

\begin{document}

\title{Random phase approximation to neutrino energy losses from a
relativistic electron-positron plasma.}
\author{L. B. Leinson$^{1,2}$ and A. Pérez$²$ \\
$^{1}$Institute of Terrestrial Magnetism, Ionosphere and\\
Radio Wave Propagation\\
RAS, 142190 Troitsk, Moscow Region, Russia\\
$^{2}$Departamento de Física Teórica and IFIC, \\
Universidad de Valencia-CSIC, \\
Dr. Moliner 50, 46100--Burjassot, Valencia, Spain}
\maketitle

\begin{abstract}
The process of $\nu \bar{\nu}$ radiation from a relativistic plasma of
electrons and positrons is studied within the Random Phase Approximation. 
\end{abstract}

The neutrino emission from a relativistic electron-positron plasma plays an
important role in many astrophysical scenarios, including the processes in
degenerate helium cores of red giant stars, cooling of neutron stars and
pre-white dwarf interiors. Up to now, the corresponding neutrino energy
losses were considered as a simple sum of those caused by plasmon \footnote{%
For brevity we use the term \char`\"{}plasmons\char`\"{} both for the
transverse and longitudinal eigen modes of the plasma oscillations.} decays 
\cite{ARW63}-\cite{RDP03}, electron-positron annihilation \cite{ChM60}-\cite%
{Dicus72}, neutrino photoproduction processes \cite{Dicus72}-\cite{BPS67},
and $\nu\bar{\nu}$ bremsstrahlung from electrons \cite{FR69}-\cite{LP01}.

Except for the plasmon decay, the above calculations have been performed
neglecting electromagnetic correlations among electrons and positrons in the
medium. The first attempt to incorporate the collective effects in the
annihilation processes was made by Braaten \cite{B92}, by including plasma
corrections to the electron dispersion relation. It was shown, however, that
the plasma effects give no noticeable modification to the neutrino
emissivity from the plasma at temperatures and densities where the
annihilation processes dominate.

It is of interest, however, to take into account electromagnetic
interactions, since the interference between some of the above particular
processes, caused by the plasma polarization, can lead to non-trivial
phenomena \cite{L99}. To incorporate the collective plasma effects, it is
convenient to use the formalism of correlation functions, instead of
considering particular neutrino emitting processes. In this case, the
differential probability of the neutrino-pair emission is given by the
imaginary part of the forward scattering amplitude of the neutrino pair, as
shown by the following diagram

\psfig{file=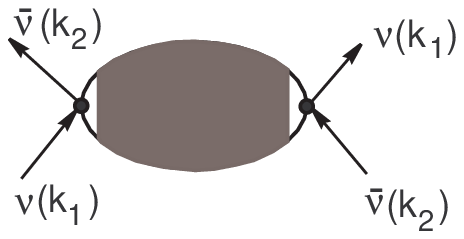}

This diagram is the lowest order in the weak interaction, but the internal
(electron) part represents the sum of all polarization graphs, which begin
and end at the weak vertices, and include all possible electromagnetic
interactions inside.

In what follows we use the Standard Model of weak interactions, the system
of units $\hbar=c=1$ and the Boltzmann constant $k_{B}=1$. The
fine-structure constant is $\alpha=e^{2}/4\pi=1/137$.

For the electron energies under consideration, the in-vacuum weak
interaction of electrons with the neutrino field can be written, in a
point-like current-current approach, as $\mathcal{H}_{eff}=-\left(G_{F}/%
\sqrt{2}\right)j^{\mu}J_{\mu}$, where $G_{F}$ is the Fermi coupling
constant, and $j^{\mu}=\bar{\nu}$\thinspace$\gamma^{\mu}\left(1-\gamma_{5}%
\right)\,\nu$ is the neutrino current. The vacuum weak current of the
electrons is of the standard form, $J_{\mu}=\bar{\psi}\Gamma_{\mu}\psi$,
where, $\psi$ represents the electron field, and the weak electron vertex, 
\begin{equation}
\Gamma_{\mu}\equiv C_{V}\gamma_{\mu}-C_{A}\gamma_{\mu}\gamma_{5},
\label{Gam}
\end{equation}
includes the vector and axial-vector terms; $C_{V}=\frac{1}{2}%
+2\sin^{2}\theta_{W}$, $C_{A}=\frac{1}{2}$ stand for electron neutrinos,
whereas $C_{V}^{\prime}=-\frac{1}{2}+2\sin^{2}\theta_{W}$ , $C_{A}^{\prime}=-%
\frac{1}{2}$\ are to be used for muon and tau neutrinos; $\theta_{W}$ is the
Weinberg angle.

The total energy which is emitted into neutrino pairs per unit volume and
time is given by the following formula: 
\begin{equation}
Q=\frac{G_{F}^{2}}{2}\sum_{\nu}\int\;\omega\frac{2Im\left[\mathrm{Tr}%
\left(j^{\mu}j^{\nu\ast}\right)\Pi_{\mu\nu}\left(-\omega,-\mathbf{k}\right)%
\right]}{e^{\omega/T}-1}\frac{d^{3}k_{1}}{2\omega_{1}(2\pi)^{3}}\frac{%
d^{3}k_{2}}{2\omega_{2}(2\pi)^{3}},  \label{Q}
\end{equation}
where $\Pi_{\mu\nu}\left(\omega,\mathbf{k}\right)$ represents the exact
retarded weak polarization tensor of the plasma, and the integration goes
over the phase space volume of neutrinos and antineutrinos of total energy $%
\omega=\omega_{1}+\omega_{2}$ and total momentum $\mathbf{k=k}_{1}+\mathbf{k}%
_{2}$. The symbol $\sum_{\nu}$\ \ indicates that summation over the three
neutrino types has to be performed, with the corresponding values of $C_{V}$%
\ and $C_{A}$, as explained above.

One can simplify this equation by inserting $\int
d^{4}K\delta^{\left(4\right)}\left(K-k_{1}-k_{2}\right)=1$, and making use
of the Lenard's integral 
\begin{equation}
\int\frac{d^{3}k_{1}}{2\omega_{1}}\frac{d^{3}k_{2}}{2\omega_{2}}%
\delta^{\left(4\right)}\left(K-k_{1}-k_{2}\right)\mathrm{Tr}%
\left(j_{\mu}j_{\nu}^{\ast}\right)=\frac{4\pi}{3}\left(K_{\mu}K_{%
\nu}-K^{2}g_{\mu\nu}\right)\theta\left(K^{2}\right)\theta\left(\omega\right),
\end{equation}
where $\theta(x)$ is the Heaviside step function. We then obtain 
\begin{equation}
Q=\frac{G_{F}^{2}}{48\pi^{5}}\sum_{\nu}\int\omega\;\frac{\left(K_{\mu}K_{%
\nu}-K^{2}g_{\mu\nu}\right)\mathrm{Im}\Pi^{\mu\nu}}{\exp\left(\frac{\omega}{T%
}\right)-1}\theta\left(K^{2}\right)\theta\left(\omega\right)d\omega d^{3}k
\label{QQ}
\end{equation}
with $K=\left(\omega,\mathbf{k}\right)$.

The relevant input for this calculation is, thus, the weak polarization
tensor, and the desired degree of accuracy is determined by the
approximations made in evaluating this quantity. In contrast to quantum
electrodynamics, where there is only one parameter in the perturbation
series, $\alpha $ (or $e^{2}/v$, $v$ being the particle velocity), the
plasma is characterized by a few additional parameters depending on the
temperature and the density. For the problem under cosideration, the most
important parameter is the ratio of the plasma frequency to the temperature $%
\omega _{p}/T$. In the high-temperature limit, $\omega _{p}^{2}/T^{2}\ll 1$,
the plasma polarization effects can be neglected, while at moderate and low
temperatures, $\omega _{p}^{2}/T^{2}\gtrsim 1$, the plasma polarization must
be necessarily taken into account. In what follows we show that the standard
calculation of the neutrino energy losses due to $e^{+}e^{-}$ annihilation,
as described before, are valid, strictly speaking, only in the
high-temperature limit, while the intermediate and low temperatures are also
typical for applications. To include the plasma effects we use the Random
Phase Approximation (RPA) to the weak polarization tensor of the plasma. As
we will see, this approach substantially improves the energy losses due to $%
e^{+}e^{-}$ annihilation and correctly reproduces the neutrino energy losses
caused by plasmon decays, but it does not allow to describe neither the
photoproduction of neutrino pairs nor the neutrino bremsstrahlung from
electrons because the latter processes are of the next to RPA order in the
fine structure constant.

To introduce some notations to be used further, we begin with the lowest
(zero) order in $\alpha$ approximation to the polarization tensor, which is
given by the one-loop diagram:

\psfig{file=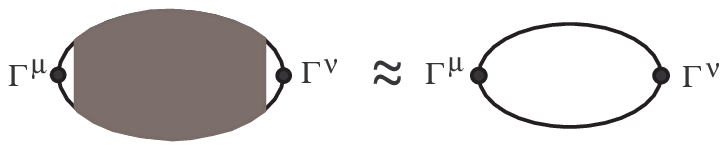}

This approach assumes that the net negative electric charge in a
hot relativistic plasma of electrons is cancelled by a uniform background of
ions.

By using the above expression for the weak vertex $\Gamma _{\mu }$ of the
electron, one can write the lowest order polarization tensor as follows: 
\begin{equation}
\Pi _{0}^{\mu \nu }\left( \omega ,\mathbf{k}\right) =C_{V}^{2}\Pi _{VV}^{\mu
\nu }\left( \omega ,\mathbf{k}\right) +C_{A}^{2}\Pi _{AA}^{\mu \nu }\left(
\omega ,\mathbf{k}\right) +2C_{V}C_{A}\Pi _{AV}^{\mu \nu }\left( \omega ,%
\mathbf{k}\right) .  \label{P0}
\end{equation}%
Here $\Pi _{VV}^{\,\mu \nu }$, $\Pi _{AV}^{\,\mu \nu }$, and $\Pi
_{AA}^{\,\mu \nu }$ are the retarded vector-vector, axial-vector, and
axial-axial one-loop polarizations, respectively, calculated for a fixed
temperature $T$ and chemical potential $\mu $ of the plasma. We use the
exact one-loop polarization functions, obtained with the Matsubara's
technique. To save place in this short letter, we do not show these
well-known expressions.

To specify the components of the polarization tensors, we select a basis
constructed from the following orthogonal four-vectors $h^{\mu}\equiv\left(%
\omega,\mathbf{k}\right)/\sqrt{K^{2}}$and $l^{\mu}\equiv\left(k,\omega%
\mathbf{n}\right)/\sqrt{K^{2}}$, where the space-like unit vector $\mathbf{%
n=k}/k$ is directed along the space component $\mathbf{k}$ of the
transferred 4-momentum $K$. Then the longitudinal (with respect to $\mathbf{%
k)}$ basis tensor can be chosen as $L^{\rho\mu}\equiv-l^{\rho}l^{\mu}$. The
transverse components of $\Pi^{\,\rho\mu}$ have a tensor structure
proportional to the tensor $T^{\rho\mu}\equiv\left(g^{\rho\mu}-h^{\rho}h^{%
\mu}-L^{\rho\mu}\right)$, where $g^{\rho\mu}$ is the signature tensor.

In this basis, the vector-vector polarization tensor has the following form 
\begin{equation}
\Pi _{VV}^{\,\rho \mu }\left( K\right) =\pi _{l}\left( K\right) L^{\rho \mu
}+\pi _{t}\left( K\right) T^{\rho \mu },  \label{Pi}
\end{equation}%
where the longitudinal polarization function is defined as $\pi _{l}=\left(
1-\omega ^{2}/k^{2}\right) \Pi _{VV}^{\,00}$ and the transverse polarization
function is found to be $\pi _{t}=\left( g_{\rho \mu }\Pi _{V}^{\,\rho \mu
}-\pi _{l}\right) /2$. The axial-vector polarization has to be an
antisymmetric tensor. In the rest frame of the plasma, it can be written as 
\begin{equation}
\Pi _{AV}^{\,\rho \mu }\left( K\right) =\Pi _{VA}^{\,\rho \mu }\left(
K\right) =\pi _{AV}\left( K\right) i\frac{K_{\lambda }}{k}\epsilon ^{\rho
\mu \lambda 0},  \label{AV}
\end{equation}%
where $\epsilon ^{\rho \mu \lambda 0}$ is the completely antisymmetric
tensor $\left( \epsilon ^{0123}=+1\right) $ and $\pi _{AV}\left( K\right) $
is the axial-vector polarization function of the medium. As for the axial
term, it must be a symmetric tensor. The most general expression for this
tensor is, therefore 
\begin{equation}
\Pi _{A}^{\mu \nu }\left( K\right) =\pi _{l}\left( K\right) L^{\mu \nu }+\pi
_{t}\left( K\right) T^{\mu \nu }+\pi _{A}\left( K\right) g^{\mu \nu }.
\label{A}
\end{equation}%
Using Eqs. (\ref{P0} - \ref{A}) one can easily obtain from Eq. (\ref{QQ})
the zero order (in $\alpha $) expression for the neutrino emissivity: 
\begin{align}
Q_{\mathrm{0}}& =-\frac{G_{F}^{2}}{48\pi ^{5}}\sum_{\nu }\int_{2m}^{\infty
}d\omega \frac{\omega }{\exp \left( \frac{\omega }{T}\right) -1}\int_{k<%
\sqrt{\omega ^{2}-4m^{2}}}d^{3}k\,\,K^{2}  \notag \\
& \times \left[ C_{V}^{2}\left( \mathrm{Im}\pi _{l}+2\mathrm{Im}\pi
_{t}\right) +C_{A}^{2}\left( \mathrm{Im}\pi _{l}+2\mathrm{Im}\pi _{t}+3%
\mathrm{Im}\pi _{A}\right) \right] .  \label{Q0}
\end{align}%
In this expression, integration goes over the domain of time-like momentum
transfer, in agreement with the total energy and momentum of the outgoing
neutrino pair. In this case, the imaginary part of the lowest-order
polarization functions is caused by the creation and annihilation of the $%
e^{+}e^{-}$ pairs in the plasma, and exists only for $K^{2}>4m^{2}$.

According to the unitarity theorem, the diagrams for any particular weak
processes, for a given approximation to the weak polarization tensor, can be
obtained by cutting the forward scattering amplitude of the neutrino pair
across the lines of the intermediate states, as shown by the dashed line:

\psfig{file=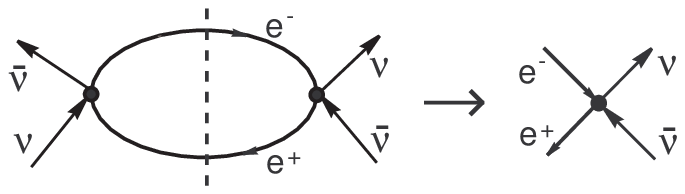}

The matrix elements obtained in this way correspond exactly to those used by
previous authors in order to calculate the neutrino energy losses due to
annihilation of $e^{+}e^{-}$ pairs. Therefore, there is no need to proceed
with the analysis of Eq. (\ref{Q0}). From the optical theorem, it is clear
that this equation yields the same neutrino emissivity obtained by Dicus 
\cite{Dicus72} with the aid of the above matrix elements\footnote{%
We have checked this explicitly.}:%
\begin{align}
Q_{\mathrm{0}}& =\frac{G_{F}^{2}m^{9}}{18\pi ^{5}}\sum_{\nu }\left[ \left(
7C_{V}^{2}-2C_{A}^{2}\right) \left(
G_{0}^{-}G_{-1/2}^{+}+G_{0}^{+}G_{-1/2}^{-}\right) \right.  \notag \\
& +9C_{V}^{2}\left( G_{0}^{-}G_{1/2}^{+}+G_{0}^{+}G_{1/2}^{-}\right) +\left(
C_{V}^{2}+C_{A}^{2}\right) \left(
4G_{1}^{-}G_{1/2}^{+}+4G_{1}^{+}G_{1/2}^{-}\right.  \notag \\
& \left. \left.
-G_{1}^{-}G_{-1/2}^{+}-G_{0}^{+}G_{1/2}^{-}-G_{0}^{-}G_{1/2}^{+}-G_{1}^{+}G_{-1/2}^{-}\right) 
\right] .  \label{Dicus}
\end{align}%
In this expression, the functions $G_{n}^{\pm }\left( \lambda ,\nu \right) $
with 
\begin{equation}
\lambda =\frac{T}{m},\,\,\,\,\,\,\nu =\frac{\mu }{T}
\end{equation}%
are defined as follows 
\begin{equation}
G_{n}^{\pm }\left( \lambda ,\nu \right) \equiv \lambda ^{3+2n}\int_{\lambda
^{-1}}^{\infty }dx\frac{x^{2n+1}\left( x^{2}-\lambda ^{-2}\right) ^{1/2}}{%
\exp \left( x\pm \nu \right) +1}.
\end{equation}

How accurate is this approach? If the polarization effects are important,
the minimal approximation to the polarization tensor for a Coulomb system of
particles requires summation of all the ring diagrams. The well known
example to this is the photon propagator in the plasma. It is represented by
an infinite sum of ring diagrams and has to be found from the Dyson's
equation, depicted graphically as

\psfig{file=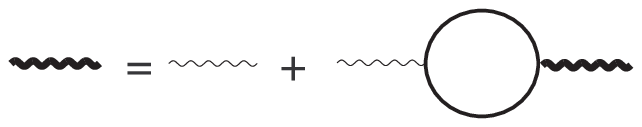}

Here, the thin dashed-line represents $\mathrm{D}_{\rho \lambda
}^{0}\left( K\right) $ -\ {}the photon propagator in vacuum. The solution to
this equation is also well known: it consists on the sum of the longitudinal
and transverse terms%
\begin{equation}
D_{\lambda \rho }\left( K\right) =D_{l}\left( K\right) L_{\lambda \rho
}+D_{t}\left( K\right) T_{\lambda \rho }  \label{D}
\end{equation}%
with 
\begin{equation}
D_{l}\left( K\right) =\frac{1}{K^{2}-e^{2}\pi _{l}\left( K\right) }%
,\,\,\,\,\,\,\,\,D_{t}\left( K\right) =\frac{1}{K^{2}-e^{2}\pi _{t}\left(
K\right) }.  \label{Dlt}
\end{equation}

\noindent With the aid of this in-medium photon propagator, the minimal
approach to the weak polarization tensor reduces to the sum of the following
two diagrams,

\psfig{file=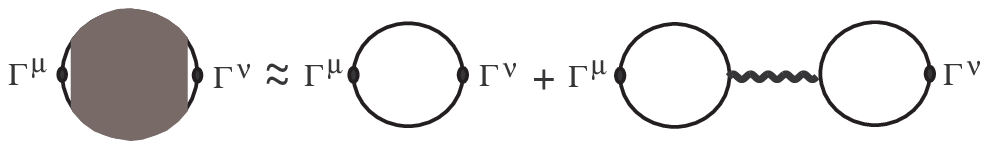}

\noindent where the thick dashed line stands for the in-medium photon
propagator $D^{\rho \lambda }\left( K\right) $, as given by Eqs. (\ref{D}), (%
\ref{Dlt}). In this way, we have collected the infinite number of ring
diagrams in the weak polarization tensor. As it is well known, this
corresponds to RPA. The irreducible polarization insertions are taken here
in the one-loop approximation, therefore this weak polarization tensor does
not allow to describe neither the photoproduction of neutrino pairs nor the
neutrino bremsstrahlung from electrons. The latter processes, caused by the
photon exchange among the in-medium particles, appear due to electromagnetic
corrections, in the irreducible polarization insertions, as shown in the
following diagramms 

\psfig{file=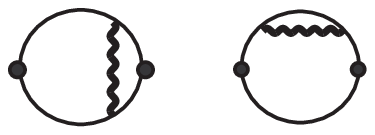}

These corrections are proportional to fine structure constant. In
contrast, the RPA corrections, described above, contribute through the
parameter $\omega _{p}/T$, and, at low temperatures, can be of the same
order as the main terms\footnote{%
Just for the same reason, the well known Eqs. (\ref{Dlt}), as obtained in
RPA, perfectly describe the plasma polarization in the photon propagator.}.
Therefore, such corrections to the irreducible polarization insertions are
out of scope of our consideration. Moreover, in the RPA, we shall omit all
extra terms which are simply proportional to $\alpha $.

Then, instead of Eq. (\ref{P0}), we have%
\begin{align}
\Pi _{\mathrm{RPA}}^{\mu \nu }& =C_{V}^{2}\left( \Pi _{VV}^{\mu \nu
}+e^{2}\Pi _{VV}^{\mu \lambda }D_{\lambda \rho }\Pi _{VV}^{\rho \nu }\right)
+C_{A}^{2}\left( \Pi _{AA}^{\mu \nu }+e^{2}\Pi _{AV}^{\mu \lambda
}D_{\lambda \rho }\Pi _{VA}^{\rho \nu }\right)  \notag \\
& +2C_{V}C_{A}\left( \Pi _{AV}^{\mu \nu }+e^{2}\Pi _{VV}^{\mu \lambda
}D_{\lambda \rho }\Pi _{VA}^{\rho \nu }\right) .  \label{RPA}
\end{align}%
In RPA we deal with both weak interactions of the electrons with the
neutrino field, and electromagnetic interactions of the electrons with the
plasma background. Therefore it is convenient to introduce short notations
for the electromagnetic polarization functions: 
\begin{equation}
\Pi _{L}=4\pi \alpha \frac{k^{2}}{\omega ^{2}-k^{2}}\pi _{l}(\omega ,k),
\label{Lem}
\end{equation}%
\begin{equation}
\Pi _{T}=4\pi \alpha \pi _{t}(\omega ,k),  \label{Tem}
\end{equation}%
\begin{equation}
\,\,\Pi _{AV}=4\pi \alpha \pi _{AV}(\omega ,k)  \label{AVem}
\end{equation}%
By inserting Eq. (\ref{RPA}) into Eq. (\ref{QQ}), and after a lengthy
(although straightforward) calculation, we obtain the RPA formula for the $%
\nu \bar{\nu}$ emissivity from the plasma, as consisting on the axial and
vector contributions: 
\begin{equation}
Q_{\mathrm{RPA}}=Q_{\mathrm{A}}+Q_{\mathrm{V}}^{\mathrm{L}}+Q_{\mathrm{V}}^{%
\mathrm{T}}+Q_{\mathrm{A}}^{\mathrm{T}},  \label{RPAtot}
\end{equation}%
where the axial term $Q_{\mathrm{A}}$ is given by the integral over the
domain $\omega ^{2}>k^{2}+4m^{2}$, compatible to the kinematics of creation
and annihilation of the $e^{+}e^{-}$ pair of total energy $\omega $ and
total momentum $\mathbf{k}$. 
\begin{eqnarray}
Q_{\mathrm{A}} &=&-\frac{G_{F}^{2}}{12\pi ^{4}}\sum_{\nu
}C_{A}^{2}\int_{2m}^{\infty }\frac{d\omega \ \omega }{\exp \left( \frac{%
\omega }{T}\right) -1}\int_{0}^{\sqrt{\omega ^{2}-4m^{2}}}dk\,\,  \notag \\
&&\times k^{2}\left( \omega ^{2}-k^{2}\right) \left( \mathrm{Im}\pi _{l}+2%
\mathrm{Im}\pi _{t}+3\mathrm{Im}\pi _{A}\right)  \label{RPA1}
\end{eqnarray}%
The remaining contribution to the neutrino energy losses consists on the
longitudinal and transverse parts, as given by the following integrals 
\begin{eqnarray}
Q_{\mathrm{V}}^{\mathrm{L}} &=&-\frac{G_{F}^{2}}{48\pi ^{5}\alpha }\sum_{\nu
}C_{V}^{2}\int_{0}^{\infty }\frac{d\omega \omega }{\exp \left( \frac{\omega 
}{T}\right) -1}\int_{0}^{\omega }dk\,\,k^{4}\left( \omega ^{2}-k^{2}\right)
^{2}  \notag \\
&&\times \frac{\mathrm{Im}\Pi _{L}}{\left( k^{2}-\mathrm{Re}\Pi _{L}\right)
^{2}+\left( \mathrm{Im}\Pi _{L}\right) ^{2}},  \label{RPAl}
\end{eqnarray}%
\begin{align}
Q_{\mathrm{V}}^{\mathrm{T}}& =-\frac{G_{F}^{2}}{24\pi ^{5}\alpha }\sum_{\nu
}C_{V}^{2}\int_{0}^{\infty }\frac{d\omega \omega }{\exp \left( \frac{\omega 
}{T}\right) -1}\int_{0}^{\omega }dkk^{2}\left( \omega ^{2}-k^{2}\right) ^{3}
\notag \\
& \times \frac{\mathrm{Im}\Pi _{T}}{\left( \omega ^{2}-k^{2}-\mathrm{Re}\Pi
_{T}\right) ^{2}+\left( \mathrm{Im}\Pi _{T}\right) ^{2}},  \label{RPAt}
\end{align}%
\begin{align}
Q_{\mathrm{A}}^{\mathrm{T}}& =-\alpha \frac{2G_{F}^{2}}{3\pi ^{3}}\sum_{\nu
}C_{A}^{2}\int_{0}^{\infty }d\omega \frac{\omega }{\exp \left( \frac{\omega 
}{T}\right) -1}\int_{0}^{\omega }dk\,\,k^{2}\left( \omega ^{2}-k^{2}\right) 
\notag \\
& \times \,\left[ \left( \left( \mathrm{Re}\pi _{AV}\right) ^{2}-\left( 
\mathrm{Im}\pi _{AV}\right) ^{2}\right) \frac{\mathrm{Im}\Pi _{T}}{\left(
\omega ^{2}-k^{2}-\mathrm{Re}\Pi _{T}\right) ^{2}+\left( \mathrm{Im}\Pi
_{T}\right) ^{2}}\right.  \notag \\
& \left. +2\mathrm{Re}\pi _{AV}\mathrm{Im}\pi _{AV}\frac{\left( \omega
^{2}-k^{2}-\mathrm{Re}\Pi _{T}\right) }{\left( \omega ^{2}-k^{2}-\mathrm{Re}%
\Pi _{T}\right) ^{2}+\left( \mathrm{Im}\Pi _{T}\right) ^{2}}\right] ,
\label{RPAA}
\end{align}%
The integrand in these contributions is proportional to the spectral
function of the in-medium longitudinal and transverse photons%
\begin{equation}
A_{L}\left( \omega ,k\right) =-\frac{1}{\pi }\frac{\mathrm{Im}\Pi _{L}}{%
\left( k^{2}+\mathrm{Re}\Pi _{L}\right) ^{2}+\left( \mathrm{Im}\Pi
_{L}\right) ^{2}},  \label{Lsf}
\end{equation}%
\begin{equation}
A_{T}\left( \omega ,k\right) =-\frac{1}{\pi }\frac{\mathrm{Im}\Pi _{T}}{%
\left( \omega ^{2}-k^{2}-\mathrm{Re}\Pi _{T}\right) ^{2}+\left( \mathrm{Im}%
\Pi _{T}\right) ^{2}},  \label{SpFun}
\end{equation}%
and the integration goes over the domain $\omega ^{2}>k^{2}$. To analyze
this part of the neutrino energy losses it is reasonable to divide this
domain of $\omega $ and $k$ into two parts. The first one corresponds to $%
0<\omega ^{2}-k^{2}<4m^{2}$, where the imaginary part of the polarizations
vanishes. In this case, the spectral functions of the in-medium photons
reduce to $\delta $-functions 
\begin{equation}
\lim_{\mathrm{Im}\Pi _{L}\rightarrow 0}A_{L}\left( \omega ,k\right) =\frac{1%
}{k^{2}}\delta \left( 1-\frac{1}{k^{2}}\Pi _{L}(\omega ,k)\right) ,
\label{LA}
\end{equation}%
\begin{equation}
\lim_{\mathrm{Im}\Pi _{T}\rightarrow 0}A_{T}\left( \omega ,k\right) =\delta
\left( \omega ^{2}-k^{2}-\Pi _{T}(\omega ,k)\right) .  \label{TA}
\end{equation}%
By performing the integral over $d\omega $, one can easily show (see the
Appendix in \cite{LP01}), that the integration over the domain $0<\omega
^{2}-k^{2}<4m^{2}$ yields the neutrino-pair emissivity due to the decay of
the transverse and longitudinal plasmons as considered in \cite{BS93}. We
shall not consider these particular terms.

Consider now contributions from the second domain, $\omega ^{2}-k^{2}>4m^{2}$%
, where the imaginary part of the polarizations does not vanish. The
axial-vector contribution (\ref{RPAA}) represents a small $\alpha $
correction, which has to be omitted since we do not include the
corresponding corrections in the irreducible polarization insertions. Thus,
in RPA, the neutrino energy losses due to the annihilation processes are the
sum of the axial and vector contributions%
\begin{equation}
Q_{\mathrm{RPA}}^{\mathrm{annih}}=Q_{\mathrm{A}}+Q_{\mathrm{V}}^{\mathrm{L}%
}+Q_{\mathrm{V}}^{\mathrm{T}}.
\end{equation}%
\begin{align}
Q_{\mathrm{RPA}}^{\mathrm{annih}}& =-\frac{G_{F}^{2}}{12\pi ^{4}}\sum_{\nu
}\int_{2m}^{\infty }d\omega \frac{\omega }{\exp \left( \frac{\omega }{T}%
\right) -1}\int_{0}^{\sqrt{\omega ^{2}-4m^{2}}}dk\,\,k^{2}\left( \omega
^{2}-k^{2}\right)  \notag \\
& \times \left[ \tilde{C}_{\mathrm{V}}^{\mathrm{L}}\left( \omega ,k\right) 
\mathrm{Im}\pi _{l}+2\tilde{C}_{\mathrm{V}}^{\mathrm{T}}\left( \omega
,k\right) \mathrm{Im}\pi _{t}+C_{A}^{2}\left( \mathrm{Im}\pi _{l}+2\mathrm{Im%
}\pi _{t}+3\mathrm{Im}\pi _{A}\right) \right] .  \label{QRPA}
\end{align}%
with 
\begin{equation}
\tilde{C}_{\mathrm{V}}^{\mathrm{L}}\left( \omega ,k\right) =\frac{k^{4}}{%
\left( k^{2}-\mathrm{Re}\Pi _{L}\right) ^{2}+\left( \mathrm{Im}\Pi
_{L}\right) ^{2}}C_{V}^{2}  \label{L}
\end{equation}%
\begin{equation}
\tilde{C}_{\mathrm{V}}^{\mathrm{T}}\left( \omega ,k\right) =\frac{\left(
\omega ^{2}-k^{2}\right) ^{2}}{\left( \omega ^{2}-k^{2}-\mathrm{Re}\Pi
_{T}\right) ^{2}+\left( \mathrm{Im}\Pi _{T}\right) ^{2}}C_{V}^{2}  \label{T}
\end{equation}

In the high-temperature regime $\omega _{p}^{2}/T^{2}\ll 1$, or in the low
density regime $\omega _{p}^{2}/m^{2}\ll 1$, when the plasma polarization
effects can be neglected, one has $\omega ^{2}-k^{2}\sim T^{2},m^{2}$ and $%
\Pi _{L,T}\sim \omega _{p}^{2}$. Eq.(\ref{QRPA}) and Eqs. (\ref{L})-(\ref{T}%
) completely reproduce the one-loop energy losses, as given by Eq.(\ref{Q0}%
). Indeed,\ in the above limits, we have 
\begin{equation*}
\lim_{\omega _{p}^{2}/m^{2}\rightarrow 0}\tilde{C}_{\mathrm{V}}^{\mathrm{L}%
}\left( \omega ,k\right) =\lim_{\omega _{p}^{2}/m^{2}\rightarrow 0}\tilde{C}%
_{\mathrm{V}}^{\mathrm{T}}\left( \omega ,k\right) =C_{V}^{2}.
\end{equation*}

We have considered the Random Phase Approximation to neutrino energy losses
from a relativistic electron-positron plasma. The main results of our
approach are Eqs. (\ref{QRPA}), (\ref{L}) and (\ref{T}), representing the
corrected neutrino energy losses due to annihilation of electrons and
positrons. The well-known energy losses due to the decay of real (on-shell)
photons are also well reproduced in the RPA.

Although the RPA is a good approximation to the neutrino energy losses due
to annihilation of electrons and positrons, and reproduces accurately the
neutrino energy losses caused by the plasmon decays, it is not yet
sufficient for a complete description of the neutrino energy losses from a
hot plasma. Indeed, the RPA does not allow to describe, for example, the
photoproduction of neutrino pairs, nor the neutrino bremsstrahlung from
electrons, which require the next (in $\alpha $) approach to the weak
polarization tensor of the medium.

\textbf{Acknowledgments}

This work has been supported by Spanish Grants AYA2004-08067-C01,
FPA2005-00711 and GV2005-264.

\end{document}